\documentclass[a4paper,12pt]{article}
\usepackage{amsfonts,amsthm,amsmath,amssymb,graphicx,hyperref,youngtab,color,tcolorbox}
\numberwithin{equation}{section}
\usepackage{bbold}
\usepackage{amsmath}
\usepackage{bm}
\begin{document}
\newtheorem{definition}{Definition}[section]
\newcommand{\be}{\begin{equation}}
\newcommand{\ee}{\end{equation}}
\newcommand{\bea}{\begin{eqnarray}}
\newcommand{\eea}{\end{eqnarray}}
\newcommand{\LE}{\left[}
\newcommand{\me}{\mathrm{e}}
\newcommand{\R}{\right]}
\newcommand{\nn}{\nonumber}
\newcommand{\Tr}{\text{Tr}}
\newcommand{\N}{\mathcal{N}}
\newcommand{\G}{\Gamma}
\newcommand{\vf}{\varphi}
\newcommand{\LL}{\mathcal{L}}
\newcommand{\Op}{\mathcal{O}}
\newcommand{\HH}{\mathcal{H}}
\newcommand{\arctanh}{\text{arctanh}}
\newcommand{\up}{\uparrow}
\newcommand{\down}{\downarrow}
\newcommand{\ket}[1]{\left| #1 \right>}
\newcommand{\bra}[1]{\left< #1 \right|}
\newcommand{\ketbra}[1]{\left|#1\right>\left<#1\right|}
\newcommand{\rd}{\partial}
\newcommand{\de}{\partial}
\newcommand{\ba}{\begin{eqnarray}}
\newcommand{\ea}{\end{eqnarray}}
\newcommand{\db}{\bar{\partial}}
\newcommand{\we}{\wedge}
\newcommand{\ca}{\mathcal}
\newcommand{\lr}{\leftrightarrow}
\newcommand{\f}{\frac}
\newcommand{\s}{\sqrt}
\newcommand{\vp}{\varphi}
\newcommand{\hvp}{\hat{\varphi}}
\newcommand{\tvp}{\tilde{\varphi}}
\newcommand{\tp}{\tilde{\phi}}
\newcommand{\ti}{\tilde}
\newcommand{\ap}{\alpha}
\newcommand{\pr}{\propto}
\newcommand{\mb}{\mathbf}
\newcommand{\ddd}{\cdot\cdot\cdot}
\newcommand{\no}{\nonumber \\}
\newcommand{\la}{\langle}
\newcommand{\lb}{\rangle}
\newcommand{\ep}{\epsilon}
 \def\we{\wedge}
 \def\lr{\leftrightarrow}
 \def\f {\frac}
 \def\ti{\tilde}
 \def\ap{\alpha}
 \def\pr{\propto}
 \def\mb{\mathbf}
 \def\ddd{\cdot\cdot\cdot}
 \def\no{\nonumber \\}
 \def\la{\langle}
 \def\lb{\rangle}
 \def\ep{\epsilon}

\title{\bf Invariant Operators, Orthogonal Bases and Correlators in General Tensor Models}
\author{\textsc{Pablo Diaz\thanks{pablodiazbe@gmail.com} \,\,\, Soo-Jong Rey\thanks{sjrey@snu.ac.kr}} \\
\\
{\small \emph{$^{*}$Fields, Gravity \& Strings, CTPU,}} 
{\small \emph{Institute for Basic Science, Daejeon 34126 \rm KOREA}}\\
{\small \emph{$^*$Department of Physics \& Astronomy, University of Lethbridge, %}}\\
%{\small \emph{4401 University Drive, Lethbridge, 
Alberta, T1K 3M4 \rm CANADA}}\\
{\small \emph{$^{\dag}$School of Physics \& Astronomy, Seoul National University, Seoul 08826 \rm KOREA}}
}
\maketitle
\begin{abstract}
We study invariant operators in general tensor models. We show that representation theory provides an efficient framework to count and classify invariants in tensor models of (gauge) symmetry $G _d = U(N_1) \otimes \cdots \otimes U(N_d)$. As a continuation and completion of our earlier work, we present two natural ways of counting invariants, one for arbitrary $G_d$ and another valid for large rank of $G_d$. We construct bases of invariant operators based on the counting, and compute correlators of their elements. The basis associated with finite rank of $G_d$ diagonalizes the two-point function of the free theory. It is analogous to the restricted Schur basis used in matrix models. We show that the constructions get almost identical as we swap the Littlewood-Richardson numbers in multi-matrix models with Kronecker coefficients in general tensor models. We explore the parallelism between matrix model and tensor model in depth from the perspective of representation theory and comment on several ideas for future investigation.
\\
%\textbf{Keywords}: Tensor models, representation theory, invariants, Kronecker coefficients, orthogonal bases.
\end{abstract}
\newpage
\tableofcontents
\section{Introduction}
Tensor models, whose elementary building block consists of tensorial objects, provide a natural generalization of matrix models. 
In theoretical physics, there are various motivations that make the tensor model an interesting system to study. In one corner, the motivation comes from a scheme for studying quantum entanglement. From the quantum mechanical point of view, $d$-rank tensor models are associated with the multi-linear symmetry group $G_d({\bf N}) = U(N_1)\otimes U(N_2) \otimes \cdots \otimes U(N_d)$ acting on a tensor product Hilbert space $\mathcal{H}=\mathcal{H}_{N_1}\otimes \cdots\otimes \mathcal{H}_{N_d}$. We know that the Hilbert space of a composed physical system is the tensor product of its constituents, and quantum correlation among them is an essential aspect of entanglement in quantum mechanics \cite{EPR}. So tensor models naturally describe composite systems. Moreover, gauge invariant operators built out of tensors separate the entangled and untangled states of $\mathcal{H}$, so they can be viewed as a probe of quantum entanglement measurements \cite{CM}.

In another corner, tensor models provide a suitable scheme for studying quantum gravity. Inspired by the success of matrix models in describing two-dimensional quantum gravity \cite{earlierTM}, tensor model was proposed as a framework for describing higher-dimensional random geometry \cite{random1,random2,random3}.
Colored tensor models \cite{color1,color2} and the development of their $1/N$-expansion \cite{1N1,1N2,1N3} have triggered an upsurge of the subject and a fast growth in recent years. The introduction of color has served to overcome several difficulties that the earlier tensor models had in describing quantum gravity at dimensions greater than two. More recently, the colored tensor model have  been found in direct connection with the $\text{AdS}_2/\text{CFT}_1$ holography, as an alternative formulation of the Sachdev-Ye-Kitaev (SYK) model \cite{SYK1,SYK2,SYK3,SYK4,SYK5,SYK6,SYK7,SYK8,SYK9,SYK10} in which the necessity of quenched disorder is dispensed while exhibiting the same large-$N$ behavior \cite{Witten}, see also \cite{Razvan}. Tensor models were also studied in the non-perturbative definition of non-abelian tensor fields
\cite{Rey:2010uz}, where interesting connections with matrix factorizations and dynamical Yang-Baxter maps were found. 

The simplest yet nontrivial tensor model is the matrix model, which has been  studied extensively in the context of AdS/CFT correspondence. In the matrix model, the use of orthogonal bases for two-point functions (first for the BPS-sector \cite{CJR} and then for general bosonic sectors \cite{restricted1,restricted2,restricted3,brauer1,brauer2,Brown:2007xh,Brown:2008ij} and for including gauge field \cite{restricted4} or fermions \cite{restricted5}) (see also \cite{Azeyanagi:2017mre})) was extremely useful for computations in $\mathcal{N}=4$ super Yang-Mills theory within the so-called non-planar regime, which involves heavy operators dual to excited D-branes and macroscopic solitonic objects in the string theory side \cite{applications1,applications2,applications3,applications4,applications5}. 

In all these situations, the colored tensor model is considered as a $n$-dimensional quantum field theory (which, as originally envisioned, may eventually describe spacetime and matter in $D \ge n$ dimensions), where the fundamental degrees of freedom are tensor fields transforming as a suitable (not necessarily irreducible) representation under an internal symmetry $G_d$. While there are issues of the tensor model pertinent to the quantum field theory such as renormalizability \cite{BenGeloun:2011rc}, there are also issues associated with the internal symmetry $G_d$ that need to be understood first. These issues are largely related to the representation theory, so we will for simplicity take the colored tensor model to be zero-dimensional. The aim of this paper is to undertake detailed study of this zero-dimensional tensor model, expanding our earlier work \cite{1stpaper}. 

This paper is meant to be a comprehensive revision and completion of our earlier work \cite{1stpaper}. Thus, there is a significant overlap with the first paper. Nevertheless, the present work ties up all the loose ends of the former by adding new proofs (like eq. ({\ref{Largefiniteagree}) which shows the match between the finite and the large $N$ operator counting, or the orthogonality of the proposed operator basis in all the labels by direct computation of the correlators in eq. (\ref{correexact})), further examples and clarifications. Section  \ref{tensorvsmatrix} is also new.  

The paper is organized as follows. We first recapitulate aspects of basic representation theory relevant for analysis in later sections. We then count physical observables, viz. invariants of tensor fields, in section \ref{Counting}, following the steps of \cite{HW} and \cite{GR}. Kronecker coefficients appear naturally in the counting. We show that representation theory actually provides two natural ways of counting gauge invariant tensor operators. One is valid for arbitrary rank of the symmetry group $G_d$, while the other is only valid at large rank of $G_d$.  Both numbers agree for large rank. We show an explicit proof of it in  Eq. (\ref{Largefiniteagree}). %We then construct bases of invariants which diagonalize the two-point function and we compute exact correlators of the elements of the bases. 
%This article is devoted to the study of gauge invariants of colored tensors. Their counting, explicit construction of an orthogonal basis of invariants and coputation of correlators. The idea is to provide a framework for translating to tensor models the success of studying orthogonal basis and their correlators in matrix models. Success both in a computational sense, since they allwoed to carry out finite $N$ computations of heavy states, and conceptually, since it provided labels for solitonic states in string theory from the dual heavy operators in $\mathcal{N}=4$ SYM theory.  
 Guided by the counting, in Section \ref{basis}, we construct bases of  gauge invariant operators and propose a basis for tensor models with finite rank of the symmetry group $G_d$ that diagonalizes the free two-point function. In section \ref{corre}, we compute the correlators of its basis elements. There is a clear parallelism between the orthogonal basis we propose and the restricted Schur basis for $d$ bosonic species in multi-matrix models. Actually, expressions in both theories are very similar as we exchange Kronecker coefficients (tensor models) and Littlewood-Richardson numbers (multi-matrix models). We explore in depth this connection in section  \ref{tensorvsmatrix}. In section \ref{Conclusion}, we highlight our main results and discuss various open problems relegated for future investigation. 

%In  (\cite{GR}) and (\cite{HW}) the authors count the gauge invariant tensors that can be constructed with $n$ copies of a $d$-rank covariant tensor and its contravariant counterpart. They consider the cases where the tensors are colorful and colorless. In the latter case indeces of different positions (colors) in the tensor are allowed to be contracted. In section 7 of (\cite{GR}) they show how to calculate correlators of those invariants in the free theory. They do it for color tensors but an analogous treatment would follow for colorless ones. \\
%One should think of the gauge invariant objects built on $n$ copies of a $d$-rank tensor as vectors in a vector space where the free two-point function acts as a non-degenerate inner product. The challenge is to find orthogonal bases of tensor invariants under this inner product. It is an algebraic problem. The way we could try is to form linear combinations of tensor invariants weighed with some class functions of the symmetric group, in a way that when performing Wick contractions the orthogonal properties of characters (via Schur lemma) trnaslate into the orthogonality of the invariants under the two-point function. This is the way it was done for matrices. I think it may work for tensors.
%%%%%%%%%%%%%%%%%%%%%%%%%%%%%%%%%%%%%%%
\section{Setup of general tensor model}\label{preli}
We start by setting our notation and reviewing some essential facts of elementary representation theory which will be relevant throughout this work.
\paragraph{Colored tensors and gauge invariant operators} \hfill\break
Colored tensors are tensors with no further tensorial symmetry assumed. We denote a rank-$d$ covariant colored tensor as 
\begin{equation}\label{dtensor}
 \Phi=\Phi_{i_1i_2\dots i_d}~e^{i_1}\otimes e^{i_2}\otimes \cdots \otimes e^{i_d},
\end{equation}
 where $\{e^{i_k}, i_k=1,\dots, N_k\} $ are complex-valued unit vectors forming an orthonormal basis of the vector space $\mathbb{C}^{N_k}$. The tensor components  $ \Phi_{i_1i_2\dots i_d}$ transform covariantly under the action of \begin{equation}\label{gauge group}
 G_d :=U(N_1)\otimes U(N_2)\otimes \cdots \otimes U(N_d)
\end{equation}
according to
\begin{equation}\label{unitaryaction}
 \Phi_{j_1j_2\dots j_d}=\sum_{i_1,\dots, i_d}U(N_1)_{j_1}^{i_1}U(N_2)_{j_2}^{i_2} \cdots U(N_d)_{j_d}^{i_d}  \Phi_{i_1i_2\dots i_d}.
\end{equation}
The complex conjugate is a rank-$d$ contravariant tensor that transforms as
\begin{equation}\label{unitaryactionconjugate}
 \overline{\Phi}^{j_1j_2\dots j_d}=\sum_{i_1,\dots, i_d}\overline{U}(N_1)^{j_1}_{i_1}\overline{U}(N_2)^{j_2}_{i_2} \cdots \overline{U}(N_d)^{j_d}_{i_d}  \overline{\Phi}^{i_1i_2\dots i_d}.
\end{equation}

We are interested in the $n$-fold tensor product $\Phi^{\otimes n}$, built out of $n$ copies of the tensor in Eq.(\ref{dtensor}). For these objects, we will use indices $i^p_k$ where $p=1,\dots,n$ and $k=1,\dots,d$. So, a basis of $\Phi^{\otimes n}$ can be written as
\begin{equation}
\bigotimes_{p=1}^n\bigotimes_{k=1}^d e^{i^p_k}\qquad \mbox{where} \qquad i^p_k=1,\dots, N_k.
\end{equation}
%Let us call $V_n$ the vector space spanned by $\text{Sym}(\Phi)^{\otimes n}$, and $\bar{V}_n$ its dual space. The vector space $V_n$ is isomorphic to $\text{Sym}(\otimes_{i=1}^d \mathbb{C}^{N_i})^{\otimes n}$ and the elements of a basis can be written as
%\begin{equation}
%(e^{i^1_1}\otimes e^{i^1_2}\otimes \cdots \otimes e^{i^1_d})\odot  (e^{i^2_1}\otimes e^{i^2_2}\otimes \cdots \otimes e^{i^2_d})\odot  \cdots \odot  (e^{i^n_1}\otimes e^{i^n_2}\otimes \cdots \otimes e^{i^n_d}).
%\end{equation}
%We will use the string of indices
%\begin{equation}
%I_k=(i^1_k,\dots,i^n_k)\longrightarrow e^{I_k}\equiv e^{i^1_k}\otimes e^{i^2_k}\otimes \cdots \otimes e^{i^n_k},
%\end{equation}
%and use the notation for permutations
%\begin{equation}
%\alpha(I_k)=(i_k^{\alpha(1)},\dots, i_k^{\alpha(n)}), \quad \alpha\in S_n.
%\end{equation}
%We will write
%\begin{equation}
%\Phi^{\otimes n}=\Phi_{I_1I_2\dots I_d}~e^{I_1}\otimes e^{I_2}\otimes \cdots \otimes e^{I_d},
%\end{equation}
%and analogously for the contravariant complex conjugate tensors.
Note that the group $G_d$ acts diagonally ($n$ times) on $\Phi^{\otimes n}$. Now, we want the copies to be indistinguishable, and so we will take the average $\text{Sym}(\Phi)^{\otimes n}$ whose components are
\begin{equation}\label{symmetrization}
\big[\text{Sym}(\Phi)^{\otimes n}\big]_{j_1^1\dots j_d^1j_1^2\dots j_d^2\dots \dots j_1^n\dots j_d^n }\equiv \frac{1}{n!}\sum_{\sigma\in S_n}\prod_{p=1}^n  \Phi_{j^{\sigma(p)}_1\dots j_d^{\sigma(p)}}.
\end{equation}
Let us consider, for fixed $n$, operators  of the form
\begin{equation}\label{oper}
\mathcal{O}=\text{Sym}(\Phi)^{\otimes n}\otimes\text{Sym}(\overline{\Phi})^{\otimes n},
\end{equation}
and select the subset of these operators which are invariant under the action of $G_d$. We will refer to them as $\mathcal{O}^{G_d-\text{inv}}$.

\paragraph{Partitions:} \hfill\break
Partitions of $n$ elements in $r$ parts $n=n_1+n_2+\dots +n_r$ are represented by Young diagrams with $n$ boxes and $r$ rows. To refer to Young diagrams, we will use the Greek letters $\mu,\nu, \lambda, \cdots$ throughout this paper. For the partition we have written, the size and number of rows of the Young diagram are expressed as $|\mu|=n$ and $l(\mu)=r$, respectively. Young diagrams are central objects in representation theory as labels for irreducible representations (irreps). They label the irreps of the symmetric group $S_n$ and the irreps of $U(N)$, which will be referred once and again in this paper. \\
If $\alpha$ is an element of $S_n$, we will denote by $[\alpha]$ its equivalence class by conjugation. All permutations with the same cycle structure belong to the same equivalence class. Thus, $[\alpha]$ is naturally associated with the Young diagram built on the cycle structure of $\alpha$. We will denote its size  by $|[\alpha]|$; it counts the number of elements that belong to that class. For instance, $|[1]|=1$ since the identity is the only element that belongs to its class, and the Young diagram associated with the the equivalence class of the identity is $(1^n)$, that is, the one column diagram.

\paragraph{Representation space.} \hfill\break
As we have shown in Eq.(\ref{unitaryaction}) and Eq.(\ref{unitaryactionconjugate}), $\Phi$ and $\overline{\Phi}$ admit a $G_{d}$-action. This action can be extended diagonally  to $\text{Sym}(\Phi)^{\otimes n}$ and $\text{Sym}(\overline{\Phi})^{\otimes n}$. Let us call $V_n$ and its isomorphic complex conjugate $\overline{V}_n$ the vector spaces spanned by $\text{Sym}(\Phi)^{\otimes n}$ and $\text{Sym}(\overline{\Phi})^{\otimes n}$, respectively. As representation spaces, under the action of the gauge group $G_d$, $V_n$ and $\overline{V}_n$ split into orbits that correspond to irreducible representations (irreps) of $G_d$. It is known that the irreducible representations of $G_d$  are labeled by a collection of $d$ Young diagrams with $n$ boxes each $(\mu_1,\dots,\mu_d)$ whose number of rows do not exceed the rank of their group. That is, $l(\mu_i)\leq N_i$. As these irreps will appear often along this paper, the following shorthand notation will be introduced
\begin{equation}\label{mu}
\bm{\mu}\equiv\mu_1,\dots, \mu_d.
\end{equation}
We will use this notation for expressions like $\delta_{\bm{\mu}\bm{\nu}}$, meaning that
\begin{equation}
\delta_{\bm{\mu}\bm{\nu}}\longleftrightarrow \mu_k=\nu_k, \qquad  (k=1,\dots, d).
\end{equation}

The objects $\Phi^{\otimes n}$ and $\overline{\Phi}^{\otimes n}$ also admit an action of $d$ copies of the symmetric group 
$S_n^d=S_n\times\cdots \times S_n$. An example of this action is the symmetrization (\ref{symmetrization}). Elements of $S_n^d$ are collections $(\alpha_1,\dots,\alpha_d)$, where $\alpha_i\in S_n$. For this element, the shorthand notation will be introduced
\begin{equation}
\bm{\alpha}\equiv (\alpha_1,\dots, \alpha_d).
\end{equation}  
The same notation will hold for a collection of products of permutations, and so the meaning of notations
\begin{equation}
\bm{\alpha}\bm{\beta}=(\alpha_1\beta_1,\dots, \alpha_d\beta_d),
\end{equation}
and 
\begin{equation}
\bm{\alpha^{-1}}\equiv (\alpha_1^{-1},\dots, \alpha_d^{-1})
\end{equation}  
is clear by themselves. 

The action of $S_n^d$ on $\Phi^{\otimes n}$ is defined by
\begin{equation}
\bm{\alpha}\cdot \Phi^{\otimes n}\equiv\sum_{\bm{\alpha}\in S_n^d}\prod_{p=1}^n  \Phi_{j^{\alpha_1(p)}_1\dots j_d^{\alpha_d(p)}}, \quad \bm{\alpha}\in S_n^d,
\end{equation}
and gets extended to $V_n$ by linearity. The same applies to $\overline{V}_n$.
In general, under the action of $S_n^d$, $V_n$ and $\overline{V}_n$ split into orbits labeled by irreps of $S_n^d$ which, as already discussed above Eq.(\ref{mu}), are known to be also labeled by Young diagram $\bm{\mu}$. This is a consequence of the Schur-Weyl duality, as the two actions, $S_n^d$ and $G_d$, defined above commute. As such, in what follows, operators of colored tensor fields will be labeled by irreps of the symmetric group $S_n^d$ and the gauge group $G_d$. 

\paragraph{Schur-Weyl duality:} \hfill\break
Schur-Weyl duality states that, as the action of $S_n$ and the action of diagonal $U(N)$ on $(\mathbb{C}^N)^{\otimes n}$ commute, we have the multiplicity-free product decomposition
\begin{equation}\label{Schurweylduality}
(\mathbb{C}^N)^{\otimes n}=\bigoplus_{\substack{\mu\vdash n\\
l(\mu)\leq N}}R_N(\mu)\otimes \Gamma_{\mu},
\end{equation}  
where $R_N(\mu)$ and $\Gamma_{\mu}$ are irreps of $U(N)$ and $S_n$, respectively. In the context of rank-$d$ colored tensors, the Schur-Weyl duality (\ref{Schurweylduality}) applies $d$ times, one per factor $U(N_k)$ of $G_d$ which is paired with a factor $S_n$ of $S_n^d$. It results in $\prod_{k=1}^d(\mathbb{C}^{N_k})^{\otimes n}$ decomposing in the direct sum of tensor product of representations labeled by $\bm{\mu}$. We used this extension above in labeling operators of colored tensor fields. 

\paragraph{Projectors:} \hfill\break
Given a linear sum decomposition Eq.(\ref{Schurweylduality}) according to the Schur-Weyl duality, it is possible to project into invariant subspaces labeled by $\mu$. The projectors are easily constructed from the characters of the symmetric group $S_n$ as
\begin{equation}
P^{\mu}=\frac{d_{\mu}}{n!}\sum_{\sigma\in S_n}\chi_{\mu}(\sigma)\sigma,
\end{equation}
where $d_{\mu}$ is the dimension of the irrep $\Gamma_{\mu}$, and $\chi_{\mu}$ is the character of the symmetric group associated with the partition $\mu$, which is $\chi_{\mu}(\sigma)=\text{tr}\big(\Gamma_{\mu}(\sigma)\big)$.\\
The action of the projectors on $(\mathbb{C}^{N})^{\otimes n}$ is given by
\begin{equation}
P^{\mu}(\mathbb{C}^{N})^{\otimes n}=R_N(\mu)\otimes \Gamma_{\mu}.
\end{equation}
The dimension of this subspace is $\text{Dim}_N(\mu)\,d_{\mu}$, where $\text{Dim}_N(\mu)$ is the dimension of the irrep $R_N(\mu)$ of $U(N)$.
 Now, for rank-$d$ colored tensor fields, as we applies the Schur-Weyl duality $d$ times, we find it useful to define the projectors
\begin{equation}\label{P}
\mathcal{P}^{\bm{\mu}}\equiv \frac{d_{\bm{\mu}}}{(n!)^d}\sum_{\bm{\sigma}\in S_n^d}\chi_{\mu_1}(\sigma_1)\cdots \chi_{\mu_d}(\sigma_d)\sigma_1\cdots \sigma_d,
\end{equation}
with $d_{\bm{\mu}}=d_{\mu_1}\cdots d_{\mu_d}$, and each $\sigma_k$ acting on its corresponding tensor index $i_k$. The action of $\mathcal{P}^{\bm{\mu}}$ projects  $\prod_{k=1}^d(\mathbb{C}^{N_k})^{\otimes n}$  into the isotypical space
\begin{equation}
\mathcal{P}^{\bm{\mu}}\prod_{k=1}^d(\mathbb{C}^{N_k})^{\otimes n}=R_N(\mu_1)\otimes \cdots \otimes R_N(\mu_d)\otimes \Gamma_{\mu_1}\otimes \cdots\otimes \Gamma_{\mu_d}
\end{equation}
 labeled by $\bm{\mu}=(\mu_1,\dots,\mu_d)$, what can be interpreted as a number of copies of the irrep $R_N(\mu_1)\otimes \cdots \otimes R_N(\mu_d)$ of $G_d$, each copy being labeled by a different state of the irrep $\Gamma_{\mu_1}\otimes \cdots\otimes \Gamma_{\mu_d}$ of $S_n^d$.\\
A comment is in order here. The projectors defined in \eqref{P} can be used to construct a certain set of operators explicitly, as we will do in \eqref{operatorsprojection} which are orthogonal but, in general, will not form a basis. The projectors we will actually  use to construct a basis of invariants are defined formally in \eqref{defpro}. Projectors in \eqref{defpro} act in the space where the $n$ tensor fields have been  {\it symmetrized}, implementing then the manifest symmetry of $\Phi^{\otimes n}$ under permuting any of the tensor copies. Unfortunately, we cannot give an explicit expression of the projectors \eqref{defpro}. However, their defining properties turn out to be sufficient in order to compute correlators among invariants.

\paragraph{Deltas and traces:} \hfill\break
When computing correlators of tensor observables, we find, for each tensor index $k$, products of Kronecker delta symbols
\begin{equation}\label{deltas}
\delta^{i_k^1}_{j_k^1}\cdots \delta^{i_k^n}_{j_k^n}, \qquad (i_k, j_k=1,\dots, N_k).
\end{equation}
We use properties of these expressions which emanate from the fact that the product of Kronecker deltas in Eq.(\ref{deltas}) is invariant under permuting each single Kronecker delta symbol. So, we have
\begin{equation}
\delta^{i_k^{\alpha(1)}}_{j_k^{\beta(1)}}\cdots \delta^{i_k^{\alpha(n)}}_{j_k^{\beta(n)}}=\delta^{i_k^{1}}_{j_k^{\beta\alpha^{-1}(1)}}\cdots \delta^{i_k^{n}}_{j_k^{\beta\alpha^{-1}(n)}}=\delta^{i_k^{\alpha\beta^{-1}(1)}}_{j_k^{1}}\cdots \delta^{i_k^{\alpha\beta^{-1}(n)}}_{j_k^{n}},\qquad \alpha,\beta\in S_n.
\end{equation}
We also have products of Kronecker deltas with the indices contracted. In this case, we obtain the number
\begin{equation}
N^{C(\beta)}_k=\delta^{i_k^{1}}_{i_k^{\beta(1)}}\cdots \delta^{i_k^{n}}_{i_k^{\beta(n)}},\qquad i_k=1,\dots,N_k,\quad \beta\in S_n,
\end{equation}
where $C(\beta)$ is the number of cycles of permutation $\beta$.
We also find useful the formula 
\begin{equation}\label{Ncexpansion}
N^{C(\beta)}_k=\sum_{\lambda\vdash n}\chi_{\lambda}(\beta)\text{Dim}_{N_k}(\lambda)=\frac{1}{n!}\sum_{\lambda\vdash n}d_{\lambda}\chi_{\lambda}(\beta)f_{\lambda}(N_k),
\end{equation}
where, in the last equality we have just rewrite $\text{Dim}_{N_k}(\lambda)$ using the combinatorial function 
\begin{equation}
f_{\lambda}(N_k)=\prod_{i,j}(N_k-i+j),
\end{equation}
where $i,j$ are the coordinates of the Young diagram $\lambda$ starting from the top left. So, $i$ is the row number and $j$ is the column number.\\
The formula (\ref{Ncexpansion}) originates from the relation between power sums and Schur functions in the context of symmetric functions
\begin{equation}
p_{\sigma}(x_1,\dots,x_N)=\sum_{\lambda\vdash n}\chi_{\lambda}(\sigma)s_{\lambda}(x_1,\dots,x_N),
\end{equation}
when we specialize all variables to 1 \footnote{See \cite{M} for details.}. It turns out that
\begin{equation}
p_{\sigma}(\underbrace{1,\dots,1}_N)=N^{C(\sigma)}, \quad s_{\lambda}(\underbrace{1,\dots,1}_N)=\text{Dim}_N(\lambda),
\end{equation}
so (\ref{Ncexpansion}) follows immediately. \\
Products of deltas can also be used to define ``the trace'' of a function of $S_n$ as
\begin{equation}
\text{Tr}_{V}(f)\equiv \sum_{\alpha\in S_n}f(\alpha) \delta^{i_k^{1}}_{i_k^{\alpha(1)}}\cdots \delta^{i_k^{n}}_{i_k^{\alpha(n)}},
\end{equation}
where $V$ is the vector space we are tracing over, which in the above case is $(\mathbb{C}^{N_k})^{\otimes n}$.
An example of trace is
\begin{equation}\label{trid}
\text{Tr}_{V}(\delta)=N_k^n,
\end{equation}
where $\delta(\alpha)=0$ unless $\alpha$ is the trivial permutation in which case $\delta(1)=1$. This is ``the trace of the identity'' and gives the dimension of the entire space $V$, as shown in Eq.(\ref{trid}). The functions that we use in this paper are mainly projectors. For projectors, we have
\begin{equation}
\text{Tr}_{V}(P^{\mu})=\frac{1}{n!}\sum_{\alpha\in S_n}\chi_{\mu}(\alpha) \delta^{i_k^{1}}_{i_k^{\alpha(1)}}\cdots \delta^{i_k^{n}}_{i_k^{\alpha(n)}}=\text{dim }P^{\mu}(\mathbb{C}^{N_k})^{\otimes n}=\text{Dim}_{N_k}(\mu)\, d_{\mu}.
\end{equation}

\section{Counting invariants}\label{Counting}
Let us now count tensor field invariants. We first observe that invariants of tensors under the simultaneous unitary action (\ref{unitaryaction}) and (\ref{unitaryactionconjugate}) are obtainable from contracting in all possible ways pairs of covariant and contravariant tensors. In other words, the set
\begin{equation}\label{spanset}
\Big\{\mathcal{O}_{\bm{\alpha}}=\prod_{p=1}^n\Phi_{i_1^pi_2^p\dots i_d^p}\overline{\Phi}^{i_1^{\alpha_1(p)}i_2^{\alpha_2(p)}\dots i_d^{\alpha_d(p)}}|\bm{\alpha}\in S_n^d\Big\}
\end{equation}
spans the space of invariants. This is so because the space of $U(N_k)$-invariant linear maps
\begin{equation}\label{map}
\iota: e_i\otimes e^j\to \delta_{i}^j
\end{equation}
is one-dimensional and, as we have $n$ copies of both $\Phi$ and $\overline{\Phi}$, the map (\ref{map}) can be applied to any of the permuted slots. Obviously, this holds for each tensor index, resulting in $d$ permutations of $n$ elements for an $n$-fold product of a $d$-rank tensor, as shown in the set (\ref{spanset}).

Note that, though every invariant can be expressed as a linear combination of the elements of Eq.(\ref{spanset}), the set (\ref{spanset}) does not form a basis simply because the elements are not necessarily linearly independent. Still, the first indispensable step toward further analysis is to find a method for counting the number of $n$-fold invariants of rank-$d$ tensors. Applying arguments from representation theory, we will find two natural ways of counting invariants, one that applies to arbitrary ranks $N_k$ of the constituent unitary groups and the other that holds for large ranks $N_k$, more specifically, for $N_k\geq n$ for all $k$. Both were addressed in our previous work \cite{1stpaper}, and independently in \cite{HW} and \cite{GR}. The use of double cosets for counting problems and the subsequent use of Burnside's Lemma was developed in the context of counting Feynman diagrams in a previous paper \cite{KR}. Finite $N_k$ counting for the case of 2-rank tensors appear for the first time in the Physics literature in \cite{MR}. We will study them first and use the labels of these two methods of counting and then construct the respective bases of invariants.

%%%%%%%%%%%%%%%%%%%%%%%%%%%%%%%%%%%%%%%
\subsection{Finite $N_k$}
   
As introduced above, we will refer $V_n$ and $\overline{V}_n$ to the vector spaces spanned by $\text{Sym}(\Phi)^{\otimes n}$ and $\text{Sym}(\overline{\Phi})^{\otimes n}$, respectively.
The action of the group $G_d$ on operators $\mathcal{O}$ defined in Eq. (\ref{oper}) is given by its simultaneous diagonal action on  both $\Phi^{\otimes n}$ and $\overline{\Phi}^{\otimes n}$. As mentioned above, this action will split $V_n$ and $\overline{V}_n$, which are isomorphic each other, into representations of $G_d =U(N_1)\otimes U(N_2)\otimes \cdots \otimes U(N_d)$ which are labeled by  $\bm{\mu}=(\mu_1,\dots,\mu_d)$, where $\mu_k$ are Young diagrams with $n$ boxes. The number of rows of each diagram does not exceed $N_k$, that is, $l(\mu_k)\leq N_k$. 

%Consider the index $k$ out of the $d$ indices of $\Phi$. In the $n$-fold product $\Phi^{\otimes n}$, the space associated with this index is isomorphic to $(\mathbb{C}^{N_k})^{\otimes n}$. Now, as a consequence of Schur-Weyl duality, irreducible representations of $(\mathbb{C}^{N_k})^{\otimes n}$ under the diagonal action of $U(N_k)$ are labeled by Young diagrams with $n$ boxes with at most $N_k$ rows.  Thus, the irreducible representations of $V_n$ (and of $\overline{V}_n$ by the isomorphism) under the action of  $G_d$ are labeled by collections $\bm{\mu}=(\mu_1,\dots,\mu_d)$, where $\mu_k$ are Young diagrams with $n$ boxes, denoted as $|\mu_k|=n$. The number of rows of each diagram does not exceed $N_k$, that is, $l(\mu_k)\leq N_k$.

The problem of classifying $\mathcal{O}^{G_d-\text{inv}}$, the $G_d$-invariants of $V_n\otimes \overline{V}_n$, translates into a representation theory problem since the invariants are in one-to-one correspondence with $G_d$-invariant maps $(V_n, \overline{V}_n)\to \mathbb{C}$, that is,
\begin{equation}\label{numberofinvariants} 
\text{dim}\{\mathcal{O}^{G_d-\text{inv}}\}=\text{dim}\text{ Hom}_{G_d}(V_n,\overline{V}_n),
\end{equation}
and, by Schur's Lemma, there exists one homomorphism (modulo an equivalence) every time we pair up an irrep of $V_n$ with an irrep of $\overline{V}_n$.\\

Now we should explore the shape of $V_n$, namely the number of times a given irrep of $G_d$ happens, if any. In order to implement the diagonal symmetrization of the space $V_n$ (as it is defined)  we will use the Kronecker map.  Denote $N=N_1N_2\cdots N_d$. It is clear that one can construct a map $\otimes_{i=1}^d \mathbb{C}^{N_i}\to \mathbb{C}^N$. This is the Kronecker map, and produces an embedding of the Kronecker product of matrices $\otimes_{i=1}^dU(N_i)$ into $U(N)$. The tensor $\Phi_{i_1\dots i_d}$ gets reorganized under this map as $\Phi_I$, where now $I=1,\dots,N$. So,
\begin{equation}\label{phicomp}
\Phi'^{\otimes n}=\Phi_{I_1}\otimes \Phi_{I_n}\quad I_k=1,\dots, N,
\end{equation}
where the prime indicates that the Kronecker map has been performed. The diagonal action of the group $S_n$ on $\Phi'^{\otimes n}$ is obvious in \eqref{phicomp}. The vector space spanned by $\Phi'^{\otimes n}$ is clearly isomorphic to $(\mathbb{C}^{N})^{\otimes n}$ and so, by Schur-Weyl 
duality, we can write
\begin{equation}\label{SWphi}
\Phi'^{\otimes n}\cong \bigoplus_{\lambda\vdash n} R^N_\lambda \otimes \Gamma_\lambda,
\end{equation}
where $ R^N_\lambda$ is an irrep of $U(N)$ and $\Gamma_\lambda$ and irrep of $S_n$.
 Symmetrization of $\Phi'^{\otimes n}$ is nothing but the projection of $\Phi'^{\otimes n}$ onto the trivial representation of the diagonal action. That is, acting with $\mathcal{P}^{(n)}$ on  \eqref{SWphi}. This produces
\begin{equation}\label{Vndis}
\text{Sym}(\Phi'^{\otimes n})\cong  R^N_{(n)} \otimes \Gamma_{(n)}.
\end{equation}
The trivial representation $\Gamma_{(n)}$ is 1. So, we could actually remove it from equation \eqref{Vndis}.
 However, it is convenient to keep track of it. \\
% In turn, this maps 
%\begin{equation}
%V_n\to R_{(n)}^N,
%\end{equation}
%as $R_{(n)}^N\cong \text{Sym}(\mathbb{C}^{N})^{\otimes n}$ from the Schur-Weyl duality\footnote{The Schur-Weyl duality asserts that $(\mathbb{C}^{N})^{\otimes n}=\oplus_\lambda R^N_\lambda \otimes \Gamma_\lambda$ under the action of $U(N)$ and $S_n$, where $R^N_\lambda$ and $\Gamma_\lambda$ are irreps of $U(N)$ and $S_n$, respectively. The operation ``Sym'' projects the direct sum into the subspace labeled by $\lambda=(n)$. As $\Gamma_{(n)}$ is one-dimensional, it follows that $R_{(n)}^N\cong \text{Sym}(\mathbb{C}^{N})^{\otimes n}$.}. 
We now need to restrict to the original group $G_d$, under which the tensors transform. Indeed,
\begin{equation}
V_n=\text{Sym}(\Phi'^{\otimes n})\big\downarrow^{U(N)}_{G_d},
\end{equation}
where the restriction of the unitary groups can be seen as inverting the Kronecker map and restoring the original $d$ indices of $\Phi$ transforming under $G_d$.
The decomposition of a general irrep $R^N_{\lambda}$ of $U(N)$ when we restrict to $G_d\subset U(N)$ is known. The branching coefficients of this restriction are the Kronecker coefficients. Thus, for $|\lambda|=n$, one has
\begin{equation}\label{Kroneckerunitary}
R_{\lambda}^N\big\downarrow^{U(N)}_{G_d}=\bigoplus_{\substack{|\mu_1|,\dots,|\mu_d|=n\\
l(\mu_k)\leq N_k}}g_{\bm{\mu},\lambda}R_{\mu_1}^{N_1}\otimes \cdots \otimes R_{\mu_d}^{N_d},
\end{equation}
where $g_{\bm{\mu},\lambda}$ are the Kronecker coefficients. For the case of our interest, $\lambda=(n)$. Now, $g_{\bm{\mu},(n)}=g_{\bm{\mu}}$, as can be checked by the general formula
\begin{equation}\label{Kroneckergral}
g_{\bm{\mu}}=\frac{1}{n!}\sum_{\alpha\in S_n}\chi_{\mu_1}(\alpha)\cdots \chi_{\mu_d}(\alpha),\qquad \mu_1,\dots,\mu_d\vdash n,
\end{equation}
since $\chi_{(n)}(\alpha)=1$. So,
\begin{equation}
R_{(n)}^N\big\downarrow^{U(N)}_{G_d}=\bigoplus_{\substack{|\mu_1|,\dots,|\mu_d|=n\\
l(\mu_k)\leq N_k}}g_{\bm{\mu}}R_{\mu_1}^{N_1}\otimes \cdots \otimes R_{\mu_d}^{N_d}.
\end{equation}
We thus found the decomposition
\begin{eqnarray}\label{decomposition}
V_n \ &\cong&\bigoplus_{\substack{|\mu_1|,\dots,|\mu_d|=n\\
l(\mu_k)\leq N_k}}g_{\bm{\mu}}\Gamma_{(n)}\otimes R^{N_1}_{\mu_1}\otimes\cdots\otimes R^{N_d}_{\mu_d},\nonumber \\
&& \nonumber \\
\overline{V}_n&\cong&\bigoplus_{\substack{|\mu_1|,\dots,|\mu_d|=n\\
l(\mu_k)\leq N_k}}g_{\bm{\mu}}\overline{\Gamma}_{(n)}\otimes \overline{R}^{N_1}_{\mu_1}\otimes\cdots\otimes \overline{R}^{N_d}_{\mu_d},
\end{eqnarray}
where the representation $\overline{R}_{\mu_k}$ is isomorphic to the irrep $R_{\mu_k}$ in the contravariant basis.  %The Kronecker coefficients $g_{\bm{\mu}}$ are thus the multiplicity of irrep  $\bm{\mu}$ in the decomposition. Equivalently, $g_{\bm{\mu}}$ is the number of orbits labeled by $\bm{\mu}$ that appear in $V_n$ when acted on by $G_d$. 
In equation \eqref{decomposition} we can see that $g_{\bm{\mu}}$  counts copies of the trivial representation of the diagonal action of $S_n$, $\Gamma_{(n)}$, in the subspace   $R^{N_1}_{\mu_1}\otimes\cdots\otimes R^{N_d}_{\mu_d}$ of $V_n$. %This becomes clear by taking dimensions in \eqref{Vndis}, where $\text{dim }\text{Sym}(\Phi'^{\otimes n})=\text{dim }V_n$ can be seen as counting the number of copies of $\Gamma_{(n)}$ in $V_n$. Now, take dimensions in \eqref{decomposition} and the interpretation of the Kronecker coefficients just mentioned follows.

We can then apply the decomposition (\ref{decomposition}) into Eq.(\ref{numberofinvariants}) and obtain the formula for the dimension of distinct invariants 
\begin{tcolorbox}
\begin{equation}\label{numberofinvariantscomplete} 
\text{dim}\{\mathcal{O}^{G_d-\text{inv}}\}=\text{dim}\text{ Hom}_{G_d}(V_n,\overline{V}_n)=\sum_{\substack{|\mu_1|,\dots,|\mu_d|=n\\
l(\mu_k)\leq N_k}}g^2_{\bm{\mu}}.
\end{equation}
\end{tcolorbox}
\noindent This counting formula agrees with the result found in \cite{HW,MR,1stpaper,Robert,GR2}. Although, in this work we do not deal with fermionic fields, it is worth mentioning that the counting procedure applies identically in the fermionic case, except for the fact that the projection of the diagonal action of $S_n$ is on the sign representation, since fermion fields anti-commute. This affects the final counting: one of the $d$ partitions of the Kronecker coefficients ``absorbs'' the sign representation and gets transposed (it does not matter which one). See details in section 2.2 of \cite{Robert}.

In the table (\ref{table}), we illustrate this  result by enlisting the number of invariants for smaller values of $n$ and $N_1=N_2=N_3\equiv N$, for the case $d=3$. It illustrates rapid growth of the number of invariants as $n$ and $N$ becomes larger.

\begin{equation}\label{table}
\begin{tabular}{|c|c|c|c|c|c|}
\hline
& $N$=1&$N$=2&$N$=3&$N$=4&$N$=5 \\
\hline
$n$=1&~1&1&1&1&1\\
\hline
$n$=2&~1&4&4&4&4\\
\hline
$n$=3&~1&5&11&11&11\\
\hline
$n$=4&~1&12&31&43&43\\
\hline
$n$=5&~1&15&92&143&161\\
\hline
\end{tabular}
\end{equation}

\subsection{Large $N_k$}
If  $N_k$ were large enough, viz. $N_k\geq n$ for all $k$, there exists an alternative way of counting invariants \cite{1stpaper}, based on the observation that all invariants are expressible as linear combinations of elements in the set (\ref{spanset}), subject to equivalence of a double diagonal action of $S_n$. This is so because the initial ordering of the $n$ slots in $\Phi^{\otimes n}$ and in $\overline{\Phi}^{\otimes n}$ is irrelevant after symmetrizing. So, the number of invariants coincides with the size of double coset
\begin{equation}\label{dc}
\text{Diag}(S_n)\backslash S_n^{\times d} /\text{Diag}(S_n).
\end{equation}
The size of double coset (\ref{dc}) can be calculated using Burnside's Lemma \cite{GR,HW}. It results in the simple formula
\begin{tcolorbox}
\begin{equation}\label{numberofinvariants2}
\text{dim}\{\mathcal{O}^{G_d-\text{Inv}}\}=|\text{Diag}(S_n)\backslash S_n^{\times d} /\text{Diag}(S_n)|=\sum_{\lambda\vdash n}z_{\lambda}^{d-2}.
\end{equation}
\end{tcolorbox}
\noindent
Here,  $z_\lambda$ is the order of the centralizer of any element with cycle structure $\lambda$, which is a combinatorial number that depends on the partition $\lambda$ of $n$ as follows. If we write the partition $\lambda=(\lambda_1,\dots,\lambda_n)$ such that $n=\sum_i i\lambda_i$, then
\begin{equation}
z_\lambda=\prod_{i=1}^n i^{\lambda_i}(\lambda_i!).
\end{equation}
The number $z_\lambda$ is then related to the size of conjugacy classes by
\begin{equation}\label{ZC}
|[\alpha]|=\frac{n!}{z_{[\alpha]}},\quad \alpha\in S_n.
\end{equation} 
The formula (\ref{numberofinvariants2}) is much simpler than the formula (\ref{numberofinvariantscomplete}). Actually, computing Eq.(\ref{numberofinvariantscomplete}) rapidly becomes out of reach as $n$ grows, since there is no combinatorial method available to date for computing Kronecker coefficients. For those ranges both methods can be used, one can readily check that both formula agree each other. For instance, evaluating Eq. (\ref{numberofinvariants2}) for $d=3$ and $n=1,2,3,4,5$, we get $1,4,11,43,161$. We see that they match with the last column of Table (\ref{table}). 

In fact, it is not difficult to prove the equivalence of both formula for large ranks of $G_d$. Using the well-known orthogonality property of characters
\begin{equation}
\frac{1}{n!}\sum_{\mu\vdash n}\chi_\mu(\alpha)\chi_\mu(\beta)=\delta_{[\alpha][\beta]}z_{[\alpha]}
\end{equation}
 and the relation (\ref{ZC}), 
we see that for $N_k\geq n$,
\begin{eqnarray}\label{Largefiniteagree}
\sum_{|\mu_1|,\dots,|\mu_d|=n
}g^2_{\bm{\mu}}&=&\frac{1}{n!^2}\sum_{\alpha,\beta\in S_n}\sum_{|\mu_1|,\dots,|\mu_d|=n} \chi_{\mu_1}(\alpha)\cdots \chi_{\mu_d}(\alpha)\chi_{\mu_1}(\beta)\cdots \chi_{\mu_d}(\beta)\nonumber \\
&& \nonumber \\
&=&\frac{1}{n!^2}\sum_{\alpha,\beta\in S_n}\delta_{[\alpha][\beta]}z_{[\alpha]}^d=\sum_{[\alpha]\vdash n}\frac{|[\alpha]|^2}{n!^2} z_{[\alpha]}^d=\sum_{[\alpha]\vdash n}z_{[\alpha]}^{d-2},
\end{eqnarray}
what proves the large $N_k$ equality. 

Conceptually, this match is a consequence of the Schur-Weyl duality, which ensures that Kronecker coefficients also appear in the Kronecker product of irreps of $S_n$ as
\begin{equation}\label{Kroneckersymmetric}
\Gamma_{\mu_1}\otimes \cdots \otimes\Gamma_{\mu_d}=\bigoplus_{\mu}g_{\bm{\mu},\mu}\Gamma_{\mu},
\end{equation}
where no restriction in the number of columns of the diagrams appears. As a consequence, the formula derived from the double coset \eqref{dc} counts the number of invariants only for large $N_k$. Otherwise, this formula overestimates it.

%%%%%%%%%%%%%%%%%%%%%%%%%%%%%%%%%%%%%%%
\section{Bases of Invariant Operators}\label{basis}
Having obtained counting methods, we next move to construct explicit bases of the invariants. Still, the counting methods we developed in the previous section will serve as a guidance  for the construction. We will see that, associated with the two ``natural'' counting methods we introduced, it is possible to construct two ``natural" types of bases.

\subsection{Large $N_k$}
A basis of invariant operators can be constructed in the case that $N_k \geq n$ for all $k$. In the spirit of the double coset counting, two invariant operators $\mathcal{O}_{\bm{\alpha}}$ and $\mathcal{O}_{\bm{\beta}}$ are linearly independent if and only if there does not exist $\tau,\sigma\in S_n$ such that $\tau\alpha_i\sigma=\beta_i$ for all $i$ \footnote{Note that this condition does not guarantee linear independence if $n>N_k$ for any $k$.}. Now, for every monomial $\mathcal{O}_{\bm{\alpha}}$, we can choose a representative multiplying all the permutations by $\alpha_d^{-1}$. So, after reordering, we are left with a collection of operators
\begin{equation}
\{\mathcal{O}_{\beta_1\dots\beta_{d-1}1}|\beta_1,\dots,\beta_{d-1}\in S_n\}.
\end{equation}
These operators still have the equivalence
\begin{equation}
\mathcal{O}_{\beta_1\dots\beta_{d-1}1}\quad \sim \quad \mathcal{O}_{\tau\beta_1\tau^{-1}\dots\tau\beta_{d-1}\tau^{-1}1},
\end{equation}
otherwise, they are linearly independent. Now we choose representatives of the orbits of $(\beta_1,\dots\beta_{d-1})$ generated by simultaneous conjugation. Each representative will be a collection $(\sigma_1,\dots,\sigma_{d-1})$. Then, the set of invariants
\begin{tcolorbox}
\begin{equation}\label{basislargeN}
\{\mathcal{O}_{\sigma_1\dots\sigma_{d-1}1}|(\sigma_1,\dots,\sigma_{d-1}) \text{  representative}\}
\end{equation}
\end{tcolorbox}
\noindent forms a basis\footnote{This basis has been recently consider in \cite{Robert} under the name of ``trace basis''.}.

Recall that the set \eqref{spanset} spans the space of invariants but does not form a basis since it is over-complete. The set of operators \eqref{basislargeN} contains only representatives of the double coset \eqref{dc} so, it is a basis for large $N_k$. However, the basis \eqref{basislargeN} is not orthogonal with respect to the inner product defined by the two-point function. So, it will only have a limited utility for computations. A clear advantage of providing an orthogonal basis with simplifying expressions for the correlators is that it serves to compute correlators of generic observables, as they can always be decomposed into linear combinations of the elements of the  basis. 

\subsection{Finite $N_k$}
  Let us consider the case of finite $N_k$. As we shall see in the next section, the basis we are constructing below is indeed orthogonal, that is, it diagonalizes the two-point function. The relevant formula for the finite rank case is Eq.(\ref{numberofinvariantscomplete}). From this formula, we learn two things:
\begin{itemize}
\item[i)] The first equality of Eq.(\ref{numberofinvariantscomplete}) tells us that there exists one invariant operator every time we couple an irrep of $V_n$ with its dual in $\overline{V}_n$. If we associate each irrep of $V_n$ with a vector, then invariants are in one-to-one correspondence with vectors in the subspace of $V_n$ where there is no multiplicity. In the subspaces for which a certain irrep occurs more than once, invariants are in one-to-one correspondence with endomorphisms. For example, if a certain irrep occurs twice, there are four ways of pairing: $\{(v_1,\overline{v}_1),(v_1,\overline{v}_2),(v_2,\overline{v}_1),(v_2,\overline{v}_2)\}$.
\item[ii)] The second equality of Eq.(\ref{numberofinvariantscomplete}) tells us precise information about the decomposition of $V_n$ and the suitable labels to describe it. As can be read from of Eq.(\ref{numberofinvariantscomplete}), the set of labels that exhausts the counting is $\{\bm{\mu},ij\}$, where $\mu_k\vdash n$ with $l(\mu_k)\leq N_k$, and  $i,j=1,\dots,g_{\bm{\mu}}$. 
\end{itemize}
%As a basis of invariant operators for finite $N_k$, we propose
%\begin{equation}\label{operatorsfiniteN}
%\mathcal{O}_{\mu_1\dots\mu_d,ij}=\text{Tr}\big(V_n\mathcal{P}_{\mu_1\dots\mu_d,ij}\overline{V}_n\big),
%\end{equation}
%where $\text{Tr}$ is an instruction to contract all the tensor indices of the elements of $V_n$ with those of $\overline{V}_n$ such that the result is an invariant. Here, $\mathcal{P}_{\mu_1\dots\mu_d,ij}$ is the projector that acts on the vector space $V_n$ and projects onto the subspace labeled by $\mu_1\dots\mu_d$ (which has multiplicity $g_{\mu_1\dots\mu_d}$). As a basis of endomorphisms, we choose intertwiners labeled by $i,j$. So\footnote{Note the similarity of the basis so constructed with the restricted Schur basis built on matrix models \cite{restricted1,restricted2,restricted3}.},
%\begin{eqnarray}
%\mathcal{P}_{\mu_1\dots\mu_d,ij}\mathcal{P}_{\mu'_1\dots\mu'_d,i'j'}&=&\delta_{\mu_1\mu'_1}\cdots\delta_{\mu_d\mu'_d}\delta_{ji'}
%\mathcal{P}_{\mu_1\dots\mu_d,ij'}\nonumber \\
%\sum_{\mu_1\dots\mu_d}\sum_{i=1}^{g_{\mu_1\dots\mu_d}}\mathcal{P}_{\mu_1\dots\mu_d,ii}&=&\mathbb{1}.
%\label{above}
%\end{eqnarray}
In view of the decomposition Eq.(\ref{decomposition}), we propose the operator basis for finite $N_k$ as%\footnote{We first proposed this basis in \cite{1stpaper}. Subsequently, two equivalent descriptions using also group theoretical methods have appeared in the literature: \cite{Robert}, which exercises also fermionic fields into consideration, and  \cite{GR2} (see also \cite{Itoyama:2017wjb}). In \cite{GR2},  the elements of the basis of observables for finite rank are corresponded to the sub-algebras associated with equivalence classes of the group algebra of permutations $\mathbb{C}(S_n)$. In that framework, the Kronecker coefficients that count the multiplicities appear as the Clebsch-Gordan coefficients of symmetric groups. Both descriptions are equivalent via Schur-Weyl duality.  }
\begin{tcolorbox}
\vskip-0.5cm
\begin{eqnarray}\label{operatorsfiniteNdouble}
&& \mathcal{O}_{\bm{\mu},ij}=\text{Tr}\big(\Phi_{\bm{\mu},i}\overline{\Phi}_{\bm{\mu},j}\big),\nonumber \\
&& \overline{\mathcal{O}}_{\bm{\mu},ij}=\text{Tr}\big(\Phi_{\bm{\mu},j}\overline{\Phi}_{\bm{\mu},i}\big),
\end{eqnarray}
\end{tcolorbox}
\noindent where we have referred to $\Phi_{\bm{\mu},i}$ and $\overline{\Phi}_{\bm{\mu},j}$ for the subspaces of $V_n$ and $\overline{V}_n$ corresponding to copy $i$ and copy $j$, respectively, of the irrep labeled by   $\bm{\mu}$. In (\ref{operatorsfiniteNdouble}), $\text{``Tr''}$ is an instruction to contract all the tensor indices of the elements of $V_n$ with those of $\overline{V}_n$ such that the result is an invariant.
Remember that, from the decomposition \eqref{decomposition}, we have $i,j=1,\dots, g_{\bm{\mu}}$. As explained before, those latin indices label copies of the trivial representation of the diagonal action of $S_n$ in $V_n$, and in $\overline{V}_n$. Remember that the trivial representation of the diagonal action of the symmetric group (the symmetrization of $\Phi^{\otimes n}$) appears because the field we are considering, $\Phi$, is bosonic, and so the composite operators are invariant under permutations of the fields. That is why the operators where symmetrized. Had the operator been fermionic, and so anti-commuting, the composites would have had to be anti-symmetrized or, in other words, projected onto the diagonal action of the sign representation of $S_n$. That would affect the counting and the construction of the operators. See a detailed analysis of the fermionic case in \cite{Robert}.

A remark is in order. Recently, two equivalent descriptions using also group theoretical methods have appeared in the literature: \cite{Robert}, which takes also fermionic fields into consideration, and  \cite{GR2} (see also \cite{Itoyama:2017wjb}). In \cite{GR2},  the elements of the basis of observables for finite rank are corresponded to the sub-algebras associated with equivalence classes of the group algebra of permutations $\mathbb{C}(S_n)$. In that framework, the Kronecker coefficients that count the multiplicities appear as the Clebsch-Gordan coefficients of symmetric groups. Both descriptions are equivalent via Schur-Weyl duality.

\subsubsection*{Definition of invariants via projectors}\label{projectorsinter}

The subspaces $\Phi_{\bm{\mu},i}$ and $\overline{\Phi}_{\bm{\mu},j}$ of $V_n$ and $\overline{V}_n$, respectively, can be constructed by means of projectors $\mathcal{P}_{\bm{\mu},i}$ and $\mathcal{P}_{\bm{\mu},j}$. Although we are not able at this stage to provide an explicit construction of these projectors in terms of symmetric functions, we can define them formally and list some of their natural properties which will be useful later, when calculating correlators of invariants. 

A general covariant operator built upon $n$ symmetrized copies of $\Phi$ can be written using a generic function $f(\bm{\alpha})$ as 
\begin{equation}
\mathcal{O}_f\equiv\frac{1}{n!}\sum_{\sigma \in S_n}\sum_{\bm{\alpha} \in S_n^d}
f(\bm{\alpha})\prod_{p=1}^n\Phi_{i_1^{\alpha_1\sigma (p)}\dots i_d^{\alpha_d\sigma (p)}}=\sum_{\bm{\alpha}\in S_n^d}
f_S(\bm{\alpha})\prod_{p=1}^n\Phi_{i_1^{\alpha_1(p)}\dots i_d^{\alpha_d(p)}},
\end{equation}
where, in the second equality, we have defined the symmetrized function
\begin{equation}
f_S(\bm{\alpha})=\frac{1}{n!}\sum_{\sigma\in S_n}f(\alpha_1\sigma,\dots,\alpha_d\sigma).
\end{equation}
For a general contravariant operator, we have
\begin{equation}
\overline{\mathcal{O}}_f\equiv \sum_{\bm{\alpha}\in S_n^d}
f_S(\bm{\alpha})\prod_{p=1}^n\overline{\Phi}^{j_1^{\alpha_1(p)}\dots j_d^{\alpha_d(p)}}.
\end{equation}
Now,
the operator $\Phi_{\bm{\mu},i}$  is  the one corresponding to the specific subspace of $V_n$ labeled by $(\bm{\mu},i)$ which, as said above, is obtained by projection on $V_n$. So, let us write
\begin{equation}\label{defpro}
\Phi_{\bm{\mu},i}=\sum_{\bm{\alpha}\in S_n^d}
\mathcal{P}_{\bm{\mu},i}(\bm{\alpha})\prod_{p=1}^n\Phi_{i_1^{\alpha_1(p)}\dots i_d^{\alpha_d(p)}}.
\end{equation} 
Now, since 
\begin{equation}
\prod_{p=1}^n\Phi_{i_1^{\alpha_1(p)}\dots i_d^{\alpha_d(p)}}\overline{\Phi}^{j_1^{\beta_1(p)}\dots j_d^{\beta_d(p)}}=\prod_{p=1}^n\Phi_{i_1^{\alpha_1\beta_1^{-1}(p)}\dots i_d^{\alpha_d\beta_d^{-1}(p)}}\overline{\Phi}^{j_1^p\dots j_d^p}=\mathcal{O}_{\bm{\alpha}\bm{\beta^{-1}}},
\end{equation}
we can write our gauge invariant operators as 
\begin{equation}\label{OfromP}
 \mathcal{O}_{\bm{\mu},ij}=\text{Tr}\big(\Phi_{\bm{\mu},i}\overline{\Phi}_{\bm{\mu},j}\big)=\sum_{\bm{\alpha},\bm{\beta}\in S_n^d}\mathcal{P}_{\bm{\mu},i}(\bm{\alpha})\mathcal{P}_{\bm{\mu},j}(\bm{\alpha\beta})\mathcal{O}_{\bm{\beta}}=\sum_{\bm{\beta}\in S_n^d}\mathcal{P}_{\bm{\mu},ij}(\bm{\beta})\mathcal{O}_{\bm{\beta}}. 
\end{equation}
Here,  we have defined 
\begin{equation}
\mathcal{P}_{\bm{\mu},ij}(\bm{\beta})\equiv \sum_{\bm{\alpha}\in S_n^d}\mathcal{P}_{\bm{\mu},i}(\bm{\alpha})\mathcal{P}_{\bm{\mu},j}(\bm{\alpha\beta}),
\end{equation}\label{Pij}
\noindent which are the relevant functions for gauge invariant operators.  Note also that  the functions have already been symmetrized, that is,
\begin{equation}
\mathcal{P}_{\bm{\mu},ij}(\bm{\beta})=\frac{1}{n!}\sum_{\sigma\in S_n}\mathcal{P}_{\bm{\mu},ij}(\sigma\beta_1,\dots,\sigma\beta_d).
\end{equation}
Correlators will be computed using only properties of these composed functions. The main properties that we will use are
\begin{eqnarray}\label{prop1}
  \sum_{\bm{\alpha}\in S_n^d}\mathcal{P}_{\bm{\mu},ij}(\bm{\alpha})\mathcal{P}_{\bm{\nu},kl}(\bm{\alpha^{-1}\beta})&=&\delta_{\bm{\mu}\bm{\nu}}\delta_{jk}\mathcal{P}_{\bm{\mu},il}(\bm{\beta}), \\
\mathcal{P}_{\bm{\mu},ij}(\bm{\beta})&=&\mathcal{P}_{\bm{\mu},ji}(\bm{\beta^{-1}}) \label{inverse}
\end{eqnarray}
and the trace
\begin{equation}\label{prop2}
\text{Tr}\big(\mathcal{P}_{\bm{\mu},ij}\big)\equiv \mathcal{P}_{\bm{\mu},ij}(\bm{\alpha})\prod_{p=1}^n \delta_{i_1^{\alpha_1(p)}}^{i_1^p}\cdots \delta_{i_d^{\alpha_d(p)}}^{i_d^p}=\delta_{ij}\prod_{k=1}^d\text{Dim}_{N_k}(\mu_k). 
\end{equation}
Both Eq.(\ref{prop1}) and Eq.(\ref{prop2}) emanate from the intuitive idea of projectors and intertwiners whereas Eq.(\ref{inverse}) comes directly from the definition. Using Dirac notation the above properties for projectors become more transparent. If we associate
\begin{equation}
\mathcal{P}_{\bm{\mu},ij}\longleftrightarrow |\bm{\mu},i\rangle \langle \bm{\mu}, j|,\quad \langle \bm{\mu}, i|\bm{\nu}, j\rangle=\delta_{\bm{\mu}\bm{\nu}}\delta_{ij},
\end{equation}
 then Eq.(\ref{prop1}) and Eq.(\ref{prop2}) are obvious:
\begin{eqnarray}
 |\bm{\mu},i\rangle \langle \bm{\mu}, j |\bm{\nu},k\rangle \langle \bm{\nu}, l|&=&\delta_{jk}\delta_{\bm{\mu}\bm{\nu}} |\bm{\mu},i\rangle \langle \bm{\mu}, l|\nonumber \\
\text{Tr}\big(|\bm{\mu},i\rangle \langle \bm{\mu}, j|\big)&\propto&\delta_{ij},
\end{eqnarray}
where the proportionality in the second equation is precisely the dimension of the subspace $(\bm{\mu},i)$ as in Eq.(\ref{prop2}).

\paragraph{Orthogonal invariant operators} \hfill\break
%The projectors defined in  Eq.(\ref{P}) are related to Eq.(\ref{Pij}) as 
%\begin{equation}
% \mathcal{P}_{\bm{\mu}}=\sum_{i=1}^{g_{\bm{\mu}}}\mathcal{P}_{\bm{\mu},ii},
%\end{equation}
%that is, $ \mathcal{P}_{\bm{\mu}}$ projects onto the isotypical component. 
The projectors explicitly defined in \eqref{P} can be used to construct a set of orthogonal invariants. Associated with projectors $ \mathcal{P}_{\bm{\mu}}$, we  construct the operators
\begin{equation}\label{operatorsprojection}
\mathcal{O}_{\bm{\mu}}=\frac{d_{\mu_1}\cdots d_{\mu_d}}{n!^d}\sum_{\alpha_1,\dots,\alpha_d\in S_n} \chi_{\mu_1}(\alpha_1)\cdots \chi_{\mu_d}(\alpha_d)\mathcal{O}_{\bm{\alpha}},
\end{equation}
where $\mathcal{O}_{\bm{\alpha}}$'s are as in Eq.(\ref{spanset}).
In general, operators $\mathcal{O}_{\bm{\mu}}$ do not form a basis, except for special cases like $d=3$ and $n=1,2,3,4$, where there are no multiplicities and so they coincide with $\mathcal{O}_{\bm{\mu},ij}$. However, we have an explicit construction of them and, as we will shown below, we find that they form an orthogonal set in terms of inner product defined by the two-point function. 
An explicit construction of $\mathcal{O}_{\bm{\mu},ij}$ in terms of permutations must exist since, as discussed before, the set (\ref{spanset}) spans the space of invariants operators. In fact, In \cite{Robert},
 the same basis of operators is expressed in terms of branching coefficients.
%Finding it is tantamount to finding an explicit construction of $\mathcal{P}_{\bm{\mu},ij}$ in terms of functions of the symmetric group. We leave this task for a future work. 

%%%%%%%%%%%%%%%%%%%%%%%%%%%%%%%%%%%%%%%
\section{Correlators}\label{corre}
Consider a tensor model, defined by the partition function for a free theory,
\begin{equation}
Z=\int d\Phi d\overline{\Phi}e^{-\Phi \cdot \overline{\Phi}}.
\end{equation}
This sets the probability distribution function for evaluating correlators. 
Here, in the probability distribution function, the quadratic term $\Phi\overline{\Phi}$ is chosen to be the simplest 
\begin{equation}
\Phi \cdot \overline{\Phi}=\Phi_{i_1\dots i_d}\overline{\Phi}^{i_1\dots i_d},
\end{equation}
with repeated indices contracted. So, the two-point correlator of this tensor model reads
\begin{equation}
\langle \Phi_{i_1\dots i_d}\overline{\Phi}^{j_1\dots j_d}\rangle={1 \over Z} \int d\Phi d\overline{\Phi}~\Phi_{i_1\dots i_d}\overline{\Phi}^{j_1\dots j_d}e^{-\Phi\overline{\Phi}}=\delta_{i_1}^{j_1}\cdots \delta_{i_d}^{j_d}.
\end{equation}
If we have $n$ copies of $\Phi$ and $\overline{\Phi}$, then we get a sum over Wick contractions
\begin{equation}
\langle \prod_{p=1}^n\Phi_{i^p_1\dots i^p_d}\prod_{q=1}^n\overline{\Phi}^{j^q_1\dots j^q_d}\rangle=\sum_{\sigma\in S_n}\prod_{p=1}^n \delta_{i_1^p}^{j_1^{\sigma(p)}}\cdots \delta_{i_d^p}^{j_d^{\sigma(p)}}.
\end{equation}
The invariant operators we are considering here have the schematic structure $\mathcal{O}=\Phi^{\otimes n}\otimes \overline{\Phi}^{\otimes n}$. When computing correlators of the form $\langle\mathcal{O}\overline{\mathcal{O}}'\rangle$ we will consider each operator normal ordered, so that we will only allow contractions between $\Phi$'s of  $\mathcal{O}$ and $\overline{\Phi}$'s of $\overline{\mathcal{O}}'$ and between $\overline{\Phi}$'s of $\mathcal{O}$  and $\Phi$'s of $\overline{\mathcal{O}}'$. For this reason, the sum in the correlator $\langle\mathcal{O}\overline{\mathcal{O}}'\rangle$ will be the sum over Wick contractions determined by the two permutations $\sigma,\tau\in S_n$.  \\
For invariant operators of the form (\ref{spanset}), we have
\begin{eqnarray}\label{correlatorsalphas}
\langle\mathcal{O}_{\bm{\alpha}}\overline{\mathcal{O}}_{\bm{\beta}}\rangle&=&\sum_{\sigma,\tau\in S_n}\prod_{p=1}^n \delta^{i_1^p}_{i_1^{\sigma\alpha_1\tau\beta_1^{-1}(p)}}\cdots \delta^{i_d^p}_{i_d^{\sigma\alpha_d\tau\beta_d^{-1}(p)}}\nonumber \\
&& \nonumber \\
&=&\sum_{\sigma,\tau\in S_n}N_1^{C(\sigma\alpha_1\tau\beta_1^{-1})}N_2^{C(\sigma\alpha_2\tau\beta_2^{-1})}\cdots N_d^{C(\sigma\alpha_d\tau\beta_d^{-1})},
\end{eqnarray}
where $C(\sigma)$ is the number of disjoint cycles of permutation $\sigma$. We will use Eq.(\ref{correlatorsalphas}) and the explicit expansion of $N_k^{C(\tau)}$ given in (\ref{Ncexpansion}) to compute
the correlators of the bases we proposed in the previous section. 
Actually, Using Eq.(\ref{Ncexpansion}), we may write the correlators in terms of the characters of the symmetric group and functions $f_{\lambda}(N_k)$ as\footnote{The recent work \cite{MM} also derived an equivalent expression for the correlators. See also \cite{MM3,MM4,MM5}} 
\begin{equation}\label{correlatorsintermsofcharacters}
\langle\mathcal{O}_{\bm{\alpha}}\overline{\mathcal{O}}_{\bm{\beta}}\rangle=\frac{1}{n!^d}\sum_{\substack{\sigma,\tau\in S_n\\
\mu_1,\dots,\mu_d\vdash n}}\prod_{k=1}^{d}d_{\mu_k}\chi_{\mu_k}(\sigma\alpha_k\tau\beta_k^{-1})f_{\mu_i}(N_k).
\end{equation}
Now, let us first consider the bases we have proposed for large $N_k$. We will have
\begin{equation}\label{correlatorslargeN}
\langle\mathcal{O}_{\sigma_1\dots \sigma_{d-1}}\overline{\mathcal{O}}_{\overline{\sigma}_1\dots\overline{\sigma}_{d-1}}\rangle=\sum_{\sigma,\tau\in S_n}N_1^{C(\sigma\sigma_1\tau\overline{\sigma}_1^{-1})}\cdots N_{d-1}^{C(\sigma\sigma_{d-1}\tau\sigma_{d-1}^{-1})}N_d^{C(\sigma\tau)},
\end{equation}
where $(\sigma_1,\dots\sigma_{d-1})$ and $(\overline{\sigma}_1,\dots\overline{\sigma}_{d-1})$ refer to representatives of the orbits produced by simultaneous conjugation of the $d-1$ permutations. As anticipated in the previous section, the elements of this basis are not orthogonal under the free two-point function. Since Eq.(\ref{correlatorslargeN}) admits little simplification, there is not much useful information in these correlators. \\

%but before that let us compute the partial sums
%\begin{equation}\label{partialsums}
%\frac{d_{\mu_k}}{n!}\sum_{\alpha_k,\beta_k\in S_n}\chi_{\mu_k}(\alpha_k)\chi_{\nu_k}(\beta_k)N_k^{C(\sigma\alpha_k\tau\beta_k^{-1})}=\delta_{\mu_k\nu_k}\chi_{\mu_k}(\sigma\tau)f_{\mu_k}(N_k),
%\end{equation}

More interesting are the correlators of operators defined in Eq.(\ref{operatorsprojection}). For those operators, we have
\begin{equation}\label{correlatorsstart}
\langle\mathcal{O}_{\bm{\mu}}\overline{\mathcal{O}}_{\bm{\nu}}\rangle=\frac{1}{n!^{2d}}\sum_{\bm{\alpha},\bm{\beta}\in S_n^d} \prod_{k=1}^d d_{\mu_k}d_{\nu_k}\chi_{\mu_k}(\alpha_k)\chi_{\nu_k}(\beta_k)\langle\mathcal{O}_{\bm{\alpha}}\overline{\mathcal{O}}_{\bm{\beta}}\rangle \, . 
\end{equation}
Let us substitute Eq.(\ref{correlatorsintermsofcharacters}) into Eq.(\ref{correlatorsstart}). Using the orthogonality relation for characters:
\begin{equation}
\frac{1}{n!}\sum_{\sigma\in S_n}\chi_{\mu_k}(\sigma)\chi_{\nu_k}(\sigma^{-1}\tau)=\delta_{\mu_k\nu_k}\frac{1}{d_{\mu_k}}\chi_{\mu_k}(\tau)
\end{equation}
 for every $k=1,\dots,d$ in Eq.(\ref{correlatorsstart}), we get 
\begin{eqnarray}\label{correlatorsresult}
\langle\mathcal{O}_{\bm{\mu}}\overline{\mathcal{O}}_{\bm{\nu}}\rangle&=&\frac{1}{n!^d}\delta_{\bm{\mu\nu}}\prod_{k=1}^dd_{\mu_k}f_{\mu_k}(N_k)\sum_{\sigma\tau\in S_n}\chi_{\mu_k}(\sigma\tau)\nonumber \\
&=&\delta_{\bm{\mu\nu}}g_{\bm{\mu}}\frac{1}{n!^{d-2}}\prod_{k=1}^dd_{\mu_k}f_{\mu_k}(N_k)\nonumber \\
&=&\delta_{\bm{\mu\nu}}g_{\bm{\mu}}(n!)^2\prod_{k=1}^d\text{Dim}_{N_k}(\mu_k),
\end{eqnarray}
where $\text{Dim}_N(\mu)$ is the dimension of the irrep $\mu$ of $U(N)$. In these steps, we used Eq.(\ref{Kroneckergral}) and the fact that
\begin{equation}
\text{Dim}_N(\mu)=\frac{d_{\mu}f_{\mu}(N)}{n!}.
\end{equation}
The two-point correlators of the tensor model seems to be perfectly adapted to the classification of the invariants in terms of irreps of $V_n$, in the sense that these invariants are orthogonal under the correlators. These has been proven in Eq.(\ref{correlatorsresult}) at least for the subspaces labeled by $\bm{\mu}$. \\

It still needs to be proven that the basis operators $\mathcal{O}_{\bm{\mu},ij}$ are also orthogonal on the labels $i,j$. Now, since 
\begin{equation}\label{operatorsum}
\mathcal{O}_{\bm{\mu}}=\sum_i\mathcal{O}_{\bm{\mu},ii},
\end{equation}
the result Eq.(\ref{correlatorsresult}) suggests that 
\begin{equation}\label{orthogonalelements}
\langle\mathcal{O}_{\bm{\mu},ij}\overline{\mathcal{O}}_{\bm{\nu},kl}\rangle=n!^2\delta_{ik}\delta_{jl}\prod_{k=1}^d\delta_{\mu_k\nu_k}\text{Dim}_{N_k}(\mu_k).
\end{equation}
 Ortogonality in the $\bm{\mu}$ label follows from similar arguments as before. That is, if we write 
\begin{equation}
\langle\mathcal{O}_{\bm{\mu},ij}\overline{\mathcal{O}}_{\bm{\nu},kl}\rangle=\langle \text{Tr}\big(\Phi_{\bm{\mu},i}\overline{\Phi}_{\bm{\mu},j}\big)\text{Tr}
\big(\Phi_{\bm{\nu},l}\overline{\Phi}_{\bm{\nu},k}\big)\rangle,
\end{equation}
then, because of normal ordering, when we compute correlators $\langle\mathcal{O}\overline{\mathcal{O}}\rangle$, the Wick contractions work separately between the covariant part of $\mathcal{O}$ and the contravariant part of  $\overline{\mathcal{O}}$ and between the contravariant part of $\mathcal{O}$ and the covariant part of $\overline{\mathcal{O}}$. Now, those contractions are $G_d$-invariant (since they are deltas), so the pairing must be a homomorphism. Therefore $\bm{\mu}=\bm{\nu}$.

To prove orthogonality on the labels $i,j$ we will use the definition of the invariants we gave in (\ref{OfromP}) by means of projectors. 
The correlator reads
\begin{eqnarray}\label{correexact}
\langle\mathcal{O}_{\bm{\mu},ij}\overline{\mathcal{O}}_{\bm{\nu},kl}\rangle&=&\sum_{\bm{\alpha},\bm{\beta}\in S_n^d}\mathcal{P}_{\bm{\mu},ij}(\bm{\alpha})\mathcal{P}_{\bm{\nu},kl}(\bm{\beta})\langle\mathcal{O}_{\bm{\alpha}}\overline{\mathcal{O}}_{\bm{\beta}}\rangle \nonumber \\
&=&\sum_{\bm{\alpha},\bm{\beta}\in S_n^d}\mathcal{P}_{\bm{\mu},ij}(\bm{\alpha})\mathcal{P}_{\bm{\nu},kl}(\bm{\beta})\sum_{\sigma,\tau\in S_n}\prod_{p=1}^n \delta^{i_1^p}_{i_1^{\sigma\alpha_1\tau\beta_1^{-1}(p)}}\cdots \delta^{i_d^p}_{i_d^{\sigma\alpha_d\tau\beta_d^{-1}(p)}}\nonumber \\
&=& n!^2\sum_{\bm{\alpha},\bm{\beta}\in S_n^d}\mathcal{P}_{\bm{\mu},ij}(\bm{\alpha})\mathcal{P}_{\bm{\nu},lk}(\bm{\beta^{-1}})\prod_{p=1}^n \delta^{i_1^p}_{i_1^{\alpha_1\beta_1^{-1}(p)}}\cdots \delta^{i_d^p}_{i_d^{\alpha_d\beta_d^{-1}(p)}}\nonumber \\
&=& n!^2\sum_{\bm{\alpha},\bm{\beta}\in S_n^d}\mathcal{P}_{\bm{\mu},ij}(\bm{\alpha}\bm{\beta})\mathcal{P}_{\bm{\nu},lk}(\bm{\beta^{-1}})\prod_{p=1}^n \delta^{i_1^p}_{i_1^{\alpha_1(p)}}\cdots \delta^{i_d^p}_{i_d^{\alpha_d(p)}}\nonumber \\
&=&\delta_{lj}\delta_{\bm{\mu}\bm{\nu}} n!^2\sum_{\bm{\alpha}\in S_n^d}\mathcal{P}_{\bm{\mu},ik}(\bm{\alpha})\prod_{p=1}^n \delta^{i_1^p}_{i_1^{\alpha_1(p)}}\cdots \delta^{i_d^p}_{i_d^{\alpha_d(p)}}\nonumber \\
&=&\delta_{lj}\delta_{\bm{\mu}\bm{\nu}} n!^2\text{Tr}\big(\mathcal{P}_{\bm{\mu},ik}\big) \nonumber \\
&=&\delta_{ik}\delta_{lj}\delta_{\bm{\mu}\bm{\nu}} n!^2 \prod_{k=1}^d\text{Dim}_{N_k}(\mu_k),
\end{eqnarray}
what proves (\ref{orthogonalelements}).
The one-point function of operators $\mathcal{O}_{\bm{\mu},ij}$ can also be computed as 
\begin{equation}
\langle\mathcal{O}_{\bm{\mu},ij}\rangle=n! \sum_{\bm{\alpha}\in S_n^d}\mathcal{P}_{\bm{\mu},ij}(\bm{\alpha})\prod_{p=1}^n \delta_{i_1^{\alpha_1(p)}}^{i_1^p}\cdots \delta_{i_d^{\alpha_d(p)}}^{i_d^p}
=\delta_{ij}n!\prod_{k=1}^d\text{Dim}_{N_k}(\mu_k).
\end{equation}

Here we stress that we have noticed an interesting clue. The idea is that the correlators (\ref{orthogonalelements}) coincide with the correlators of the basis constructed recently in \cite{Robert}, and called Restricted Schur Basis (RSB), since it uses the same (representation theory) principles as their homologous matrix models. It indicates that the basis that we consider in this paper (which we proposed in \cite{1stpaper}) and the basis built in \cite{Robert} are actually the same.

\section{Relation between tensor models and matrix models}\label{tensorvsmatrix}
The similarity between the basis of operators (\ref{operatorsfiniteNdouble}) and the RSB for multi-matrix models is striking. So, in this section,  we will put both constructions in contact. Specifically, we will relate the basis (\ref{operatorsfiniteNdouble}) of rank-$d$ color tensor models with gauge group $U(N)^{\otimes d}$ to multi-matrix models with gauge group $U(N)$ of $d$ species transforming in the adjoint.\\

Let us review some basic features of the RBS  in multi-matrix models. We will offer first a brief description of the basis. Then, we will apply a similar logic as for tensor models in this paper to find a full explanation of its counting and their correlators by the only use of arguments of representation theory.  The point of view we offer is somewhat unconventional but serves us to establish a neat parallelism between the RSB and the orthogonal basis we have constructed for tensor models based exclusively on representation theory results.

\paragraph{Brief revision of the RSB} \hfill\break
In a multi-matrix model  with $d$ (bosonic) species $X_1,\dots,X_d$ and gauge group $U(N)$, the basic matrices $X_k$ transform in the adjoint as
\begin{equation}
(X_k)^{i'}_{j'}=U^{i'}_{i}(X_k)^i_{j}\overline{U}^{j'}_j,
\end{equation}
that is, as a pair of covariant and contravariant vectors.  The two-point function of the free theory reads
\begin{equation}
\langle (X_m)^i_j (\overline{X}_n)^k_l\rangle=\delta_{mn}\delta^i_l\delta^k_j.
\end{equation}
The operators we will consider are composites of $n$ fields
\begin{equation}\label{operatorstructure}
\mathcal{O}=X_1^{\otimes n_1}\otimes X_2^{\otimes n_2}\otimes\cdots\otimes  X_d^{\otimes n_d},
\end{equation}
where $n=n_1+\dots +n_d$. This partition of $n$ into $d$ parts can be represented by a Young diagram with $n$ boxes and $d$ rows that we will call $\lambda$.  Invariant operators are generated by contracting the covariant and contravariant indices in all possible ways. So, the operators
\begin{equation}\label{sigmaoperators}
\mathcal{O}_{\sigma}\equiv (X_1)^{i_1}_{i_{\sigma(1)}}\cdots (X_d)^{i_n}_{i_{\sigma(n)}},\quad \sigma\in S_n,
\end{equation}
span the space of gauge invariants of the tensor model. However, in general, the operators $\mathcal{O}_{\sigma}$ do not form a basis. For example, 
\begin{equation}\label{tausigmatau}
\mathcal{O}_{\sigma}\sim\mathcal{O}_{\tau\sigma\tau^{-1}},\quad \tau\in S_{\lambda},
\end{equation}
where we have defined $S_{\lambda}\equiv S_{n_1}\times\cdots \times S_{n_d}$. Actually, the symmetry (\ref{tausigmatau}) is the defining property of the composite operators we are considering, besides the rank of the gauge group $U(N)$.

 Among other possible bases \cite{brauer1,brauer2,Brown:2007xh,Brown:2008ij}, the RSB  \cite{restricted1,restricted2,restricted3} is relevant for us for its relation with the bases of invariants (\ref{operatorsfiniteNdouble}) constructed for tensor models. The operators of the RSB are defined as
\begin{equation}\label{RSB}
\mathcal{O}_{\mu,(\mu_1,\dots,\mu_d);ij}=\frac{1}{n_1!\cdots n_d!}\sum_{\sigma\in S_n}\text{Tr}\big(P_{\mu\to (\mu_1,\dots,\mu_d),ij}\Gamma_{\mu}(\sigma)\big)\mathcal{O}_{\sigma},\quad l(\mu),l(\mu_i)\leq N.
\end{equation}
where $\mu\vdash n$ and $\mu_i\vdash n_i$, so $(\mu_1,\dots,\mu_d)$ is an irrep of $S_{\lambda}$\footnote{Beware that, in order to avoid confusion, throughout this section we will not use the previous notation $\bm{\mu}$ to denote $(\mu_1,\dots,\mu_d)$ since each $\mu_k$ in this case labels an irrep of a different symmetric group.}. The restriction in the number of rows of the Young diagrams $\mu$ and $\mu_i$ comes from the double action $S_n\times U(N)$ that the operators admit. Then, by Schur-Weyl duality, the Young diagrams must represent irreps of $U(N)$ as well, so they cannot have more than $N$ rows. Correlators of the elements of the basis where calculated in \cite{restricted2}, and read
\begin{equation}\label{RSBcorrelators}
\langle \mathcal{O}_{\mu,(\mu_1,\dots,\mu_d);ij} \overline{\mathcal{O}}_{\nu,(\nu_1,\dots,\nu_d);kl}\rangle=\delta_{\mu\nu}\delta_{\mu_1\nu_1}\cdots\delta_{\mu_d\nu_d}\delta_{ik}\delta_{jl}\ \text{Dim}_N(\mu)\ \frac{d_{\mu_1}\cdots d_{\mu_d}}{n_1!\cdots n_d!}
\end{equation}

The projectors $P_{\mu\to (\mu_1,\dots,\mu_d),ij}$ act on the carrier space of the irrep $\Gamma_{\mu}$ and project into the subspaces labeled by $(\mu_1,\dots,\mu_d)$, which appear in the restriction $S_n\to S_{\lambda}$. The representations subduced in this restriction appear with multiplicities, and they are taken into account in the labels $i$ and $j$. The multiplicities in this restriction are the Littlewood-Richardson numbers
\begin{equation}\label{LRnumbers}
\Gamma_{\mu}\downarrow^{S_n}_{S_{\lambda}}=\bigoplus_{\mu_i\vdash n_i}g^{LR}_{\mu ;\mu_1,\dots\mu_d}\Gamma_{\mu_1}\otimes\cdots\otimes \Gamma_{\mu_d}.
\end{equation}
So, $i,j=1,\dots,g^{LR}_{\mu ;\mu_1,\dots\mu_d}$. 

Actually, strictly speaking, the operators $P_{\mu\to (\mu_1,\dots,\mu_d),ij}$ are not projectors  in the labels $i,j$ but {\it intertwiners}. That is,
 \begin{eqnarray}
P_{\mu\to (\mu_1,\dots,\mu_d),ij}P_{\mu'\to (\mu'_1,\dots,\mu'_d),i'j'}&=&\delta_{\mu\mu'}\delta_{\mu_1\mu'_1}\cdots\delta_{\mu_d\mu'_d}\delta_{ji'}
P_{\mu_1\dots\mu_d,ij'}\nonumber \\
\sum_{\mu_1\dots\mu_d}\sum_{i=1}^{g^{LR}_{\mu_1\dots\mu_d}}P_{\mu\to (\mu_1,\dots,\mu_d),ii}&=&\mathbb{1}_{\mu}.
\label{above}
\end{eqnarray}

The number of gauge invariant operators is the number of elements of the RSB, which in view of (\ref{RSB}) is
\begin{equation}\label{matrixcounting}
\text{dim}\{\mathcal{O}^{S_{\lambda}-\text{Inv}}\}=\sum_{\substack{\mu\vdash n\\
l(\mu)\leq N}}\sum_{\substack{\mu_i\vdash n_i\\
l(\mu_i)\leq N}}(g^{LR}_{\mu;\mu_1,\dots,\mu_d})^2.
\end{equation}
The number of gauge invariants operators (\ref{matrixcounting}) was calculated by direct evaluation of the partition function of the free theory \cite{Dolan}, see also \cite{Storm}. \\

So far we have simply described the RSB in a few steps. It is not the goal of this paper to sidestep and go deeper into their explicit construction or properties, which can be found extensively in \cite{restricted1,restricted2,restricted3,restricted4,restricted5} and the references therein. What we want in this section is to show how the salient features of the RSB, like the elements (invariant operators) and their two-point function, emanate from similar considerations of representation theory as used in tensor models in previous sections of this paper. We will establish a concrete parallel between both setups. We find this connection conceptually interesting and we believe that it can serve to incorporate well-developed techniques of RSB into the analysis of tensor models. %As an example of this advantage, at the end of this section, we are able to give an explicit construction of the tensor invariant operators inspired in the RSB invariants.\\

\paragraph{Interpretation from representation theory} \hfill\break
To start with, let us treat the covariant and contravariant space of the matrices separately. For that, we define
\begin{equation}
(X_k)^i_j\equiv (\xi_k)^i(\overline{\xi}_k)_j,
\end{equation}
and focus on either the covariant or the contravariant part. Let us call $W_n$ the space spanned by the contravariant pieces when we have $n$ fields, and $\overline{W}_n$ its isomorphic covariant space. When we have operators built on $n=n_1+\dots+n_d$ fields, like in (\ref{operatorstructure}), we can define a diagonal action of $U(N)$ on the contravariant part. This action commutes with the action of $S_n$ defined by permuting indices, as usual. But the structure of the operators (\ref{operatorstructure})  tells us that the group $S_{\lambda}$ acts naturally on the operators by permuting indices as well. By means of the permutation action of these two groups, $W_n$ splits into orbits induced by irreps of $S_n$, which are labeled by $\mu\vdash n$, and into orbits induced by irreps of $S_{\lambda}$, which are labeled by the collection $(\mu_1,\dots,\mu_d)$, where $\mu_i\vdash n_i$. These are the labels referred to the orbits driven by $S_n$ and $S_{\lambda}$.  Moreover, since both groups $S_n$ and $S_{\lambda}$ are not taken separately but we are considering the embedding\footnote{This embedding is analogous to the embedding $U(N_1)\otimes\cdots\otimes U(N_d)\hookrightarrow U(N)$ driven by the Kronecker map in tensor models.}
\begin{equation}
S_{\lambda}\hookrightarrow S_n,
\end{equation}
the operators  will  form at the {\it intersection} of the orbits within $W_n$ driven by the two groups. A different intersection piece in $W_n$ happens every time an irrep of $S_{\lambda}$ is subduced by an irrep of $S_n$. The multiplicities of the subduction are given by the Littlewood-Richardson numbers, as shown in (\ref{LRnumbers}). For the parallelism we are establishing here, the decomposition (\ref{LRnumbers}) is analogous to (\ref{Kroneckerunitary}) in tensor models. Now, the same can be done for the isomorphic covariant piece.\\
 In order for the complete operator $\mathcal{O}$ in (\ref{operatorstructure}) to be $U(N)$-invariant, all covariant indices must be contracted with contravariant ones. This provides a map between $W_n$ and $\overline{W}_n$. Moreover, this map must be invariant under the simultaneous action of $S_{\lambda}$ on $W_n$ and $\overline{W}_n$, as imposed by the  symmetry (\ref{tausigmatau}) on the resulting operators. So the map $W_n\to \overline{W}_n$ is actually a homomorphism of $S_{\lambda}$. Now, Schur Lemma also applies to finite groups, and tells us that the only non-null homomorphism between irreps of a finite group is the identity (up to equivalence) and happens between an irrep and a copy of itself.  In the end, the number of invariant operators in multi-matrix models is given by
\begin{tcolorbox}
\begin{equation}\label{numberofinvariantsmatrixcomplete} 
\text{dim}\{\mathcal{O}^{S_{\lambda}-\text{inv}}\}=\text{dim}\text{ Hom}_{S_{\lambda}}(W_n,\overline{W}_n)=\sum_{\substack{\mu_i\vdash n_i,\mu\vdash n\\
l(\mu_i),l(\mu)\leq N}}(g^{LR}_{\mu;\mu_1,\dots,\mu_d})^2.
\end{equation}
\end{tcolorbox}
Note the similarity between (\ref{numberofinvariantsmatrixcomplete}) and (\ref{numberofinvariantscomplete}).
The restriction in the number of rows in (\ref{numberofinvariantsmatrixcomplete}) comes, again, because of the $U(N)$ diagonal action. All the Young diagrams must, by Schur-Weyl duality, also label $U(N)$ irreps. For multi-matrix models there also exists a large $N$ simpler formula for counting invariants analogous to (\ref{numberofinvariants2}), which comes from the identity\footnote{See \cite{Storm} for a simple proof of this formula in the case of two different species.}
\begin{equation}
\sum_{\mu_i\vdash n_i,\mu\vdash n}(g^{LR}_{\mu;\mu_1,\dots,\mu_d})^2=\frac{1}{|S_{\lambda}|}\sum_{\tau\in S_{\lambda}}z_{[\tau]},
\end{equation}
where $[\tau]$ is the cycle structure of the permutation $\tau$ and $|S_{\lambda}|=n_1!\cdots n_d!$.\\

The orthogonality of the RSB with respect to the two point function can be explained in the same fashion as we have done for tensor models. When we have operators built on $n$ fields like in (\ref{operatorstructure}), the Wick contractions in $\langle \mathcal{O}\overline{\mathcal{O}}\rangle$  pair the contravariant pieces of $\mathcal{O}$ with the covariant pieces of $\overline{\mathcal{O}}$ and vice versa, analogously to the normal ordering defined in tensor models. In other words, using the notation of capital letters for a string of indices, like $I=i_1\dots i_n$, and making explicit the indices of operators $\mathcal{O}$ in (\ref{operatorstructure}), we have
\begin{equation}\label{twopointmatrix}
\langle \mathcal{O}^I_J\overline{\mathcal{O}}^K_L\rangle=\sum_{\tau\in S_{\lambda}}\delta^{\tau (I)}_L\delta^K_{\tau(J)},
\end{equation}
where $\tau(I)=i_{\tau(1)}\dots i_{\tau(n)}$. It turns out that, due to (\ref{above}), the {\it restricted characters} used to define the operators of the RSB have the same algebraic properties as the composed functions we are using to describe the tensor basis. In other words, if we associate
\begin{equation}
\text{Tr}\big(P_{\mu\to (\mu_1,\dots,\mu_d),ij}\Gamma_{\mu}(\sigma)\big)\longleftrightarrow \mathcal{P}_{\mu\to (\mu_1,\dots,\mu_d),ij}(\sigma),
\end{equation}
and use the properties (\ref{prop1}), (\ref{inverse}) and (\ref{prop2}) then the orthogonality of the RSB for multi-matrix models follows.  

The factor that appears on the right of Eq. (\ref{RSBcorrelators}) can be easily understood as the dimension of the subspace the restricted characters is projecting into. First realize that the space of invariant operators (\ref{operatorstructure}) is isomorphic to $(\mathbb{C}^{N})^{\otimes n}$. Schur-Weyl decomposition and projection onto irrep $\mu\vdash n$ tells us that the subspace we are considering is
\begin{equation}\label{SWprojected}
\mathcal{P}_{\mu}((\mathbb{C}^{N})^{\otimes n})\cong R_{\mu}\otimes \Gamma_{\mu}.
\end{equation}
The dimension of this subspace is $\text{Dim}_N(\mu)\cdot d_{\mu}$. In terms of irreps of $U(N)$, we can read (\ref{SWprojected}) as $\mathcal{P}_{\mu}((\mathbb{C}^{N})^{\otimes n})$ decomposing into as many irreps $R_{\mu}$ as the dimension of the irrep $\mu$ of $S_n$. That is, one irrep $R_{\mu}$ for each state of $\Gamma_{\mu}$. But we are projecting further on the subspace labeled by the irrep $\mu_1,\dots,\mu_d$ of $S_{\lambda}$ as subduced by $\mu$.
Now, the number of states is reduced from $d_{\mu}$ to $d_{\mu_1}\cdots d_{\mu_d}$. So, the dimension of the space the restricted characters projects onto is $\text{Dim}_N(\mu) d_{\mu_1}\cdots d_{\mu_d}$. As in the previous analysis for tensor models, the two point function computes the dimension of the space we are projecting on except for an extra $n_1!\cdots n_d!$ factor. This is what appears in (\ref{RSBcorrelators}).

The two physical systems we have considered show many similarities. Indeed, formulas for counting can be interchanged as we swap Kronecker coefficients in tensor models with Littlewood-Richardson numbers in the multi-matrix setup. This relation is highly non-trivial. The mathematical relation between   Kronecker coefficients and Littlewood-Richardson is mysterious and it is being exploited to unravel algorithms for the computation of some Kronecker coefficients, see for example \cite{Orellana}.

\section{Summary and future work}\label{Conclusion}
In this work, we have  used arguments from representation theory to count tensor invariants in color tensor models and to construct bases of these invariants based on the counting schemes. We found two different bases, one valid for arbitrary values of the ranks of the symmetry group, and a second basis of invariants which applies for large ranks. We show explicitly in Eq. (\ref{Largefiniteagree}) that the counting of elements of both bases agrees for large ranks. In each case, the invariants are found to be subspaces of a big representation space. We work out the finite rank case in which the subspaces associated with the invariants are found via projector/intertwiners introduced in subsection \ref{projectorsinter}. We next move to computing the correlators of the free theory for the elements of both bases. The finite rank basis is orthogonal under the two-point function of the free theory. There is an straight  analogy between the $d$-color finite rank basis constructed in this paper and  the restricted Schur basis used in multi-matrix models with $d$ species. The relevant difference is that whereas the multiplicity numbers in tensor models involve Kronecker coefficients they are given by Littlewood-Richardson numbers in the multi-matrix case. We explore in depth this connection in section \ref{tensorvsmatrix}. In order to put in contact both constructions we reinterpret the multi-matrix basis purely in terms of projectors from representation theory and we show that the similarities come from an identical (projection) method of construction. In the case of tensor models, the big space is the irreducible representation $R_{(n)}$ of the group $U(N_1\cdot N_2\cdots N_d)$ which, under restriction, splits into a direct sum of irreps of $G _d = U(N_1) \otimes \cdots \otimes U(N_d)$ with the Kronecker coefficients counting the multiplicities. For the multi-matrix case the big space is a given irrep of $S_n$ and the restriction to the subgroup $S_{\lambda}=S_{n_1}\times\cdots\times S_{n_d}$ produces a direct sum of irreps of $S_{\lambda}$ with multiplicities given by the Littlewood-Richardson numbers.

As pointed out in \cite{GR2}, the central role of representation theory and, specifically, of Young diagrams in both the multi-matrix model and the tensor model constructions may indicate that the relevant information of both theories (e.g. correlators) could emanate from statistical models of Young diagrams. The idea of this models is to upgrade the Young diagrams from labels to being fundamental objects. See   \cite{GR2} for details. We find it interesting and intend to pursue this as future line of study.

Another fascinating avenue of future research is the connection of tensor models with holography. One expects that holographic duals of tensor models exist  in a broader contexts than the SYK model  and that they are in some ways related to matrix models more than it is presently thought. In order to explore possible dual theories, we propose to examine holography at the level of the partition function first. It is noted that, at finite $N$, free tensor models have a Hagedorn growth of states which can be interpreted as having a phase transition \cite{BAT,BKMT}. The second phase appears at energies given by $n\sim N$, and both phases coexist for higher energies. Now, since matrix models (which have been proven successful for holography) have also a Hagedorn behavior, it is natural to inquire if tensor models with large but finite rank admit a dual description as some sort of brane systems whose dynamics is described by tensor fields. We conjecture that it is indeed so. Our idea is to utilize the mathematical fact that Kronecker coefficients (which are known to have a higher degree of complexity than Littlewood-Richardson numbers \cite{PP3}) are actually expressible as LR numbers for specific cases \cite{Bl,BVE}. These cases precisely label the specific states belonging to the energy regime $n\sim N$, where both phases of the tensor model start to coexist\cite{DR}. We therefore expect that finite rank tensor models offer a dual description of a brane system, at least at the some energy regime. We intend to report our progress into this direction in forthcoming \cite{DR} and future works.

%Regarding the two countings and the bases, there are two possible extensions of this work. First, it would be interesting to construct an orthogonal basis for large rank of the symmetry group, based on the counting Eq. (\ref{numberofinvariants2}) and perhaps using the arguments given below Eq. (\ref{zlambdasolutions}). Then, we should be able to compare both orthogonal bases, for finite and large ranks, and compute their correlators. Second, it would be useful to establish a rigorous proof of Eq.(\ref{orthogonalelements}) and, if possible, an explicit construction in terms of permutations of the invariants Eq.(\ref{operatorsfiniteNdouble}). All these progresses will be relegated to our forthcoming work \cite{DR}. 

%The tensor model we study here is bosonic. If we consider a fermionic tensor model, then we would make contact with the SYK alternative model proposed in \cite{Witten}. To build a fermionic  basis for finite rank, we would start with Eq.(\ref{operatorsfiniteNdouble}) and proceed in an analogous way as was done in  \cite{restricted5} in the context of matrix models. Then, we would be able to perform exact computations for heavy states in the model and compare them with their $\text{AdS}_2$ bulk counterparts \footnote{We would like to thank Robert de Mello Koch for proposing this idea.}. 

\vspace{0.2cm}

\section*{\bf Acknowledgment}

\noindent
We thank S. Das, D. Grumiller, R. Gurau, A.M. Polyakov, M. Walton and T. Wrase for useful discussions.  SJR acknowledges members of Technische Universit\"at Wien, Austria for many helpful conversations. The work of PD was supported partly by the Natural Sciences and Engineering Research Council of Canada, the University of Lethbridge, and by the Intitute for Basic Science (South Korea) IBS-R018-D2.

\end{document}